\documentclass[reprint,superscriptaddress,aps,prb,twocolumn,floatfix,nolinenumber]{revtex4-2}
\usepackage[utf8]{inputenc}
\usepackage{textcomp} % for writing µm
\usepackage{setspace}
\usepackage{amsmath}
\usepackage{breqn}
\usepackage{graphicx}
\usepackage{verbatim}
\usepackage{float}
\usepackage{amsfonts}
\usepackage{amssymb}
\usepackage{xcolor}
\usepackage{soul}
\usepackage{tabularx}
\setcitestyle{super}
\usepackage[colorlinks,linkcolor=blue,anchorcolor=blue,citecolor=blue,urlcolor=black]{hyperref}
\DeclareGraphicsExtensions{.pdf,.eps,.png,.jpg,.mps}
\usepackage{booktabs}
\usepackage{makecell}
\usepackage{threeparttable}
\usepackage{lineno}
\usepackage{mhchem}
\usepackage{amsmath,esint} % 输出数学公式
\usepackage{cuted}
\usepackage[none]{hyphenat}
\begin{document}
\sloppy{}
% \pagewiselinenumbers

\title{Recursive Inverse Design Enables Hyper-spectral Photonic Integrated Circuits}
\author{Hao He$^{1,\dagger}$, Zengji Tu$^{1,\dagger}$, Yuanlei Wang$^{2,\dagger}$, Hongyan Zhao$^{1}$, Chuangxin Feng$^{1}$, Yongzhuo Zhou$^{1}$, Yujun Chen$^{1}$, Ruoao Yang$^{1}$, Lei Zhang$^{1}$, Jianjun Wu$^{1}$, Qi-Fan Yang$^{2,3,4,\ast}$ and Lin Chang$^{1,3,\ast}$  \\
\vspace{3pt}
$^1$State Key Laboratory of Photonics and Communications, School of Electronics, Peking University, Beijing, 100871, China.\\
$^2$State Key Laboratory for Artificial Microstructure and Mesoscopic Physics, School of Physics, Peking University, Beijing, 100871, China\\
$^3$Frontiers Science Center for Nano-optoelectronics, Peking University, Beijing, 100876, China.\\
$^4$Peking University Yangtze Delta Institute of Optoelectronics, Nantong, Jiangsu, 226010, China\\
$^\dagger$These authors contributed equally to this work \\
\vspace{3pt}
Corresponding authors:
$^*$leonardoyoung@pku.edu.cn, $^*$linchang@pku.edu.cn.}

%\date{}

\maketitle

%\begin{abstract}
\noindent\textbf{Abstract} \\
\textbf{Spectrum manipulation is central to photonic systems, where advanced computing and sensing applications often demand highly complex spectral responses to achieve high throughput. Conventional methods for enhancing spectral complexity typically rely on cascading discrete photonic components, resulting in a complexity that scales only linearly with the number of components. Here, we introduce hyper-spectral photonic integrated circuits (HS-PICs), in which spectral complexity scales exponentially with the number of components. This is achieved through recursive inverse design—a system-level inverse design strategy that exploits intricate inter-component interactions as design freedoms, thereby substantially expanding the design space for spectral engineering. Using this approach, we demonstrate that even a single waveguide structure can resolve spectra with sub-picometer resolution, surpassing the performance of current state-of-the-art spectrometers. This performance bridges optical and microwave frequencies in spectral analysis, enabling simultaneous monitoring of optical and radio signals within a single device. Our work establishes a transformative framework for next-generation computing and sensing technologies.}\\
%\end{abstract}

\vspace{3pt}
\noindent\textbf{Introduction} \\
\noindent Spectral complexity is a key metric for optical systems, reflecting the number of independent, resolvable spectral features across a usable bandwidth. This parameter directly governs the maximum achievable density of spectral channels, and in turn, the computational throughput, multiplexing capability, and resolution attainable in applications such as optical computing\cite{liu2025Chipscale}, sensing\cite{yang2021Miniaturization}, and signal processing\cite{xu2022Selfcalibrating, liu2025Ultracompact}. While conventional free-space optics often requires bulky setups to achieve high spectral manipulation precision, integrated photonics provides a compact and scalable platform for realizing complex photonic integrated circuits (PICs). Over the past two decades, substantial progress has been made in this direction. For example, in communications, high-density spectral multiplexing in integrated photonics has enabled optical interconnect densities exceeding 5.3\,Tbps\,$mm^{-2}$\cite{daudlin2025Threedimensional, sun2025Edgeguided, yang2022Multidimensional}. Similarly, parallel processing of multiple wavelengths on-chip has facilitated photonic computing densities beyond 1\,TOPS\,$mm^{-2}$\cite{dong2024Partial, feldmann2021Parallel, bai2023Microcombbased}. In sensing, the availability of numerous independent spectral channels has boosted the development of computational spectroscopy\cite{yao2024Chipscale, zhang2025StrayLightFree}, with integrated spectrometers now rivaling the performance of their benchtop counterparts.  \\
\indent Despite these advances, current approaches for scaling up the complexity of PICs remain relatively rudimentary. In most implementations, the complexity is increased by cascading discrete photonic components such as Mach–Zehnder interferometers, ring resonators, and Fabry–Perot filters, where the overall spectral response results from the simple superposition of individual contributions\cite{horst2013Cascaded,liu2024Parallel,yako2023Videorate}. As a result, the total spectral complexity scales only linearly with the number of components. As the cascade depth increases, the number of elements, optical loss, and calibration overhead grow substantially, while the improvements in spectral structure become marginal. To overcome this limitation, several strategies have been proposed in recent years, including the use of disordered photonic media\cite{redding2013Compact,lee2025Reconstructive}, exploitation of chaotic dynamics\cite{guo2025Ultracompact,wang2014Computational}, and advanced tuning schemes\cite{liu2025Ultracompact, zhang2018Fully}. Although these methods can improve system performance to some extent, they often depend on stochastic or difficult-to-design paradigms, or require complex control procedures, thereby limiting their scalability and design flexibility. A comprehensive comparison of the existing methods on complex response generation is provided in the Supplementary material, Table.1.\\
\indent Here, we introduce hyper-spectral photonic integrated circuits to address this challenge. In contrast to the conventional counterpart, the complexity of HS-PICs scales exponentially with the number of components. This is enabled by our development of recursive inverse design—a first-of-its-kind, system-level inverse design methodology that performs hyper-spectral engineering in an end-to-end fashion. The approach is based on recursively partitioning the target response while synchronizing geometry design with layout optimization. By treating all inter-component interactions—such as interference and back-reflection—as design degrees of freedom, this method accesses a substantially larger design space than conventional inverse design techniques.\\
\indent Using this strategy, we show that even a single waveguide can be inversely designed to exhibit hyper-resolution performance exceeding that of state-of-the-art spectrometers. An HS-PIC comprising an array of such waveguides achieves unprecedented sub-picometer resolution over an 800\,nm bandwidth, in both cases limited only by our measurement setup. These hyper-spectral PICs create new application opportunities. We demonstrate that their exceptional spectral resolution bridges the conventional divide between optical and microwave domains: the same device reconstructs both optical spectra and microwave signals, with 125\,MHz radiofrequency resolution across more than 20\,THz of optical bandwidth. We successfully apply this capability to monitor signals from a microwave photonic radar and high-speed optical communication links. Owing to their superior spectral complexity, HS-PICs are poised to transform a broad range of applications in integrated photonics.\\
%Fig1
\begin{figure*}[ht]
\centering
\includegraphics[width = 17.8cm]{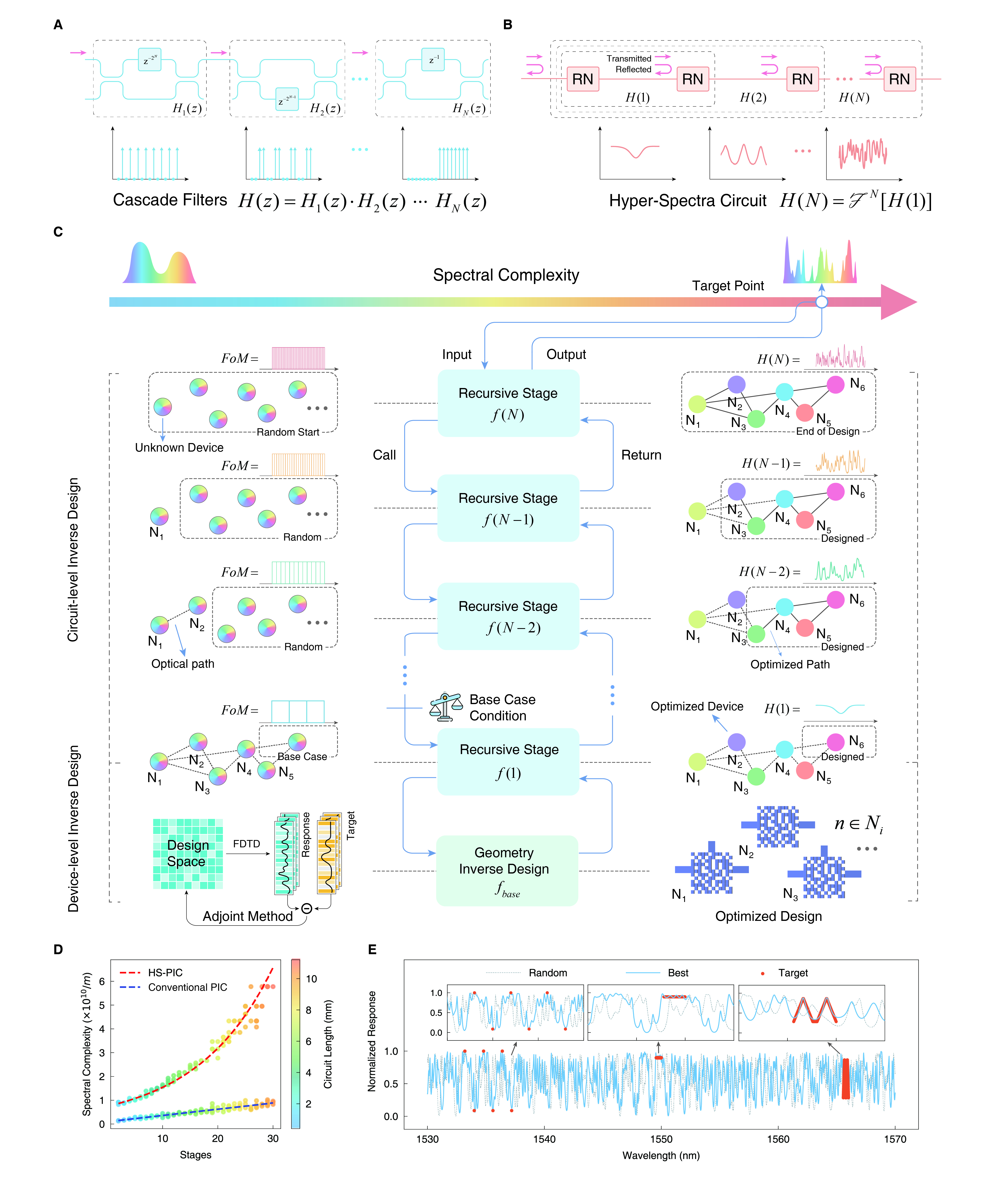}
\caption{\textbf{Concept and Principle of Hyper-spectral Circuit.} ({\bf A}) Structure of conventional optical finite impulse response (FIR) filter consists of multi-stage Mach-Zehnder interferometers. The response obtained by the multiplication of individual filters. ({\bf B}) The Structure of HS-PIC consists of arranged waveguides connected by routing node, the response is obtained by solving the recursive function $H(N)$. ({\bf C}) Schematic of the recursive inverse design. The input is the target spectrum complexity $C$. Three steps sequentially determine the node number (left), geometry (bottom), and layout (right). ({\bf D}) Simulated complexity of conventional PIC and HS-PIC with the same waveguide connection distribution. The red dashed line is the linear and exponential fits. ({\bf E}) The spectrum shaping behavior of the HS-PIC under combined targets.}  
\label{fig1}
\end{figure*}

%Result Part1
\vspace{3pt}
\noindent\textbf{The concept and principle of hyper-spectral photonic integrated circuits}\\
\noindent To compare conventional photonic integrated circuits (PIC) with the proposed hyper-spectral photonic integrated circuits, we first introduce the definition of spectral complexity. This metric characterizes the system of its maximum spectral channel density under certain independence. Mathematically, the frequency response relativity can be described as the Pearson correlation coefficient of the spectrum power transfer function $|H(\lambda)|^2$:
\begin{equation}R(\Delta \lambda)=\frac{\langle|H(\lambda)|^{2}\cdot|H(\lambda+\Delta \lambda)|^{2}\rangle_{\lambda}}{\langle|H(\lambda)|^{2}\rangle_{\lambda}\langle|H(\lambda+\Delta \lambda)|^{2}\rangle_{\lambda}}-1\end{equation}
\noindent where $H(\lambda)$, $H(\lambda+\Delta\lambda)$, and $\langle ... \rangle_\lambda$ denote the spectral response, frequency shifted response and their averaging. The correlation function measures the independence of the spectral response at neighboring wavelengths, as previous works have estimated the optical channel width with its half-width full maximum $\Delta \lambda_{0.5}$\cite{redding2013Compact}. The Spectral complexity C is then defined by generalize the reciprocal of $\lambda_{0.5}$, expressed as:
\begin{equation}C=\frac{1}{\lambda_\alpha}\end{equation}
\noindent representing the maximum spectral channel density under a specified independence constraint $\alpha$.\\
\indent For conventional photonic integrated circuits, multi-channel systems are formed by cascading filters with slightly different responses, such as ring resonators or Mach–Zehnder interferometers (MZIs)\cite{horst2013Cascaded,liu2024Parallel}. In such systems, the overall response can be expressed as the superposition of the individual component responses (Fig.~\ref{fig1}A). This process is analogous to sequential function calls in programming, leading us to hypothesize that the system exhibits a polynomial response complexity with respect to the number of components, $n$. We develop an analytical model based on unbalanced MZIs to verify this hypothesis. Through this model, we find that the system has a linear response complexity: 
\begin{equation}C_{cascade}(n)= O(n)\end{equation} 
\noindent See Supplementary Material, note 1, for details of the analytical model. Independent and identically distributed (i.i.d.) trials were performed to further validate the model. The results confirm a linearly growing trend of system complexity, revealing a fundamental limitation of conventional designs in generating ultra-complex spectra (see Supplementary Material, Method 1, for more details of the i.i.d. trials).\\
\indent To overcome this limitation and disrupt the conventional design paradigm of photonic integrated circuits, we introduce hyper-spectral photonic integrated circuits. An HS-PIC is composed of waveguide segments arranged into chain-structured or grid-structured networks, with both ends connected to a reflective routing node (RN; Fig.~\ref{fig1}B). In contrast to cascading filters that process signals sequentially, HS-PICs increase system complexity by generating circulating photonic pathways within the network. Analogous to recursive function computation, the RN reflects light and repeatedly "calls" other filters, suggesting an exponential response complexity:
\begin{equation}C_{HS-PIC}(n)= O(k^n)\end{equation} 
\noindent We quantitatively analyzed this relationship through numerical simulations of chain-structured networks with i.i.d. trials (see Supplementary Material, Method 1). The experimental results, under the same waveguide distribution as the conventional circuit, demonstrate that the new architecture exhibits a distinct exponential response complexity, enhancing that of conventional PIC by an order of magnitude within 30 devices. Notably, the system’s response also depends on waveguide connectivity. We conducted a parameter sweep over 100 HS-PIC configurations with identical routing nodes but varying waveguide lengths and computed their response complexity. Statistical evaluation shows a pronounced dependence of the response on waveguide length (see Supplementary Material, fig.\,S1). This indicates that HS-PIC introduces inter-device interactions as a new degree of freedom in the photonic design space, breaking the conventional assumption of device independence.\\
%Fig2
\begin{figure*}[ht]
\centering
\includegraphics[width = 17.8cm]{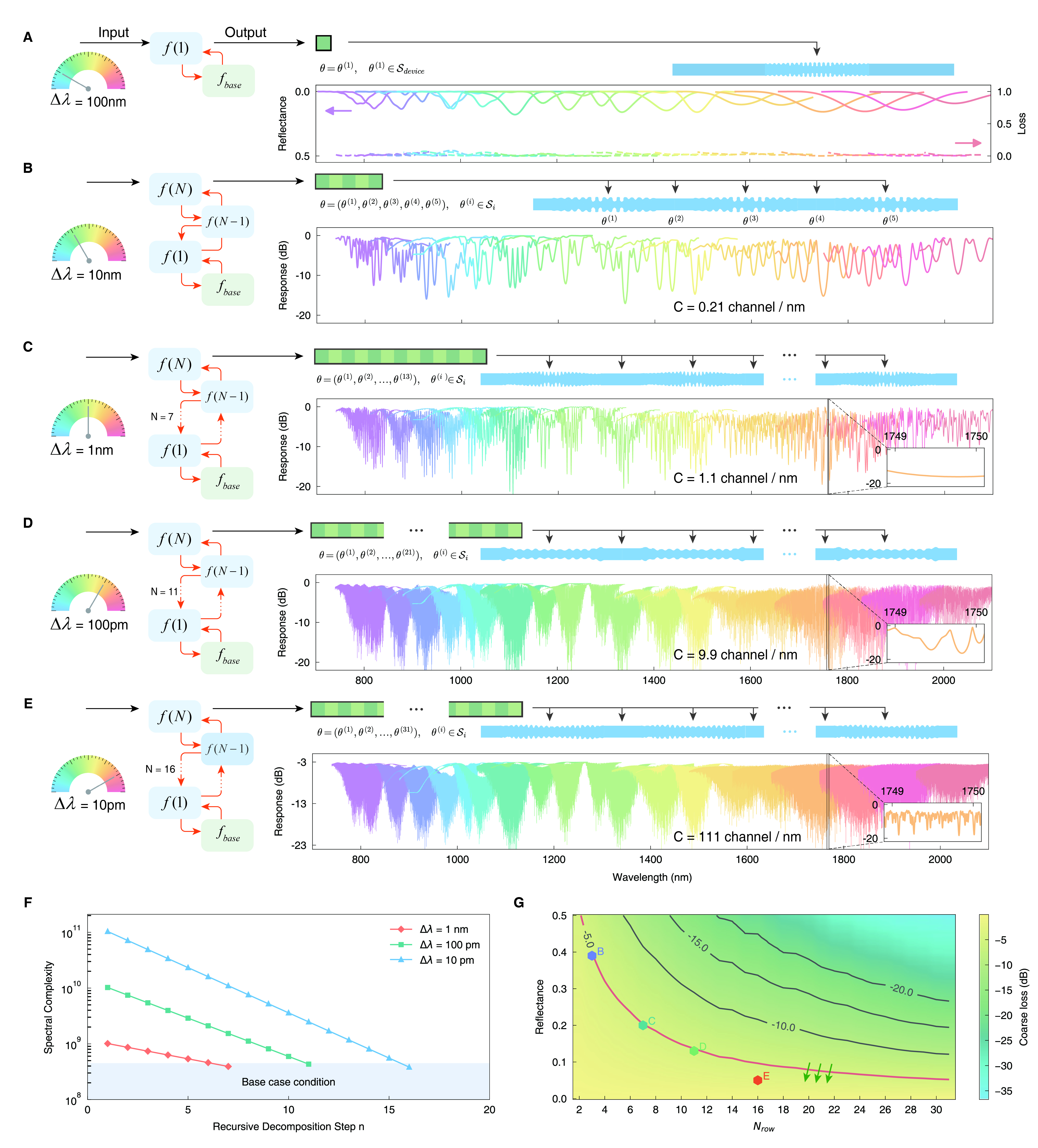}
\caption{\textbf{Recursive Inverse Design of a Straight Waveguide with Complex Response.} ({\bf A-E}) RID process with different scale complexity targets. From left to right describe the complexity input, recursive optimization, design parameters output, waveguide geometry and its response. Line colors indicate the design wavelengths.(A) One stage result is the router’s response, where dashed lines denote the insertion loss of the router, solid lines represent its reflectivity. Stage number surges from 3 to 16 with the complexity of (B) 10\,nm, (C) 1\,nm, (D) 0.1\,nm and (E) 0.01\,nm. ({\bf F}) The decomposition process of ideal chain-structured circuits, the complexity decreases as the order increases. ({\bf G}) Simulated relation of waveguide loss with respect to stage and router reflectance.}
\label{fig2}
\end{figure*}

%Result Part2
\vspace{3pt}
\noindent\textbf{Recursive inverse design for hyper-spectral engineering} \\
\noindent The HS-PIC requires the complex co-design of components and inter-component parameters, whereas conventional methodologies have typically focused on only one of these aspects at a time. Although conventional inverse-design methods can, in principle, optimize a large number of parameters simultaneously, in practice they are constrained by computational resources, limiting their application mainly to device-level problems\cite{piggott2015Inverse, molesky2018Inverse, nikkhah2024Inversedesigned, nikkhah2024Inversedesigned}. At its core, the problem requires reducing the dimensionality of a computationally intensive procedure.\\
\indent Inspired by the recursive principle in computer programming, we developed a recursive inverse-design (RID) algorithm to address this dimensionality reduction challenge. In computer science, recursion lowers computational cost by decomposing a complex task into smaller, self-similar subproblems that can be efficiently solved through repeated application of the same operation. The corresponding computational cost follows $T(n)=aT\left(\frac{n}{b}\right)+O(n^c)$, where $a<b^c$ ensures sublinear growth of $T(n)$ relative to direct computation\cite{bentley1980General}. Analogously, the recursive design framework translates the principle of recursion into the photonic design domain by breaking down a complex spectral response into hierarchically solvable subproblems. It proceeds through three sequential stages-decomposition, base-case solving, and unwinding-which collectively determine the circuit topology, device geometry, and system layout. In the decomposition stage, the target spectral complexity of order $N$ is recursively expressed as a combination of the $(N-1)$ order response and one routing element with a defined transmission matrix(Fig.~\ref{fig1}C, left). This is supported by the recursion relation derived from network simulations shown in Supplementary Material, fig.S1. When the target becomes simple enough to be solved within one connection, the base case condition is met, and recursion terminates, yielding the complete circuit architecture. Subsequently, device-level inverse design based on gradients specifies each router’s geometry within fabrication constraints, followed by binarization after continuous optimization (Fig.~\ref{fig1}C, bottom). Finally, through unwinding, global phase coherence is achieved by optimizing the interconnect lengths using particle swarm optimization, thereby aligning the composite response with the target (Fig.~\ref{fig1}C, right). The unwinding figure of merit (FoM) is defined as: 
\begin{equation}FoM=\Delta\lambda_{\alpha}+\beta\int_{\lambda}R(\Delta\lambda)\end{equation}
\noindent Here, the first term quantifies the spectral de-correlation distance, and the second term represents the frequency-integrated correlation weighted by
$\beta$. This FoM leads the interconnection towards maximum response complexity.  (see Supplementary Material, Method 2, for RID methodological details).\\
\indent We show that this recursive formulation not only accelerates convergence but also enhances scalability by performing it on idealized devices. It successfully designed distinct HS-PICs within design regions exceeding 4000\,$\mu m$ in length, whereas conventional inverse-design approaches are typically restricted to only a few tens of micrometers. The result shows that the complexity spans over two orders of magnitude variation (Fig.~\ref{fig2}F). Notably, despite the enlarged design space introduced by the recursive approach, the computational cost remains manageable. This efficiency arises from two key factors: (1) the hierarchical structure reduces high-dimensional optimization problems into simpler subproblems, and (2) a compact model library (CML) based simulation is employed for the inter-device regions, which occupy at least 95$\%$ of the chip area. All designs presented in this work were completed on dual Intel Xeon E5-2697A V4 CPUs, with total computation times under 25\,hours, confirming the practicality of the approach.\\
\indent Beyond complexity control, RID also enables spectral shaping, where the objective function is reformulated as the Euclidean distance between actual and target response:
\begin{equation}FoM=\sqrt{\sum_{n\in D}(H(n)-T(n))^{2}}\end{equation}
\noindent where $H(n)$ and $T(n)$ denote sampled response and its target defined within the domain $D$. This design approach was validated across three independent objectives and one composite objective, obtaining the results shown in Fig.~\ref{fig1}E (see Supplementary Material, Fig.\,S2, for more details regarding spectral shaping performance). Optimization under the compound objective exhibited no significant degradation compared with the single-objective cases, suggesting that the method benefits from its broad parameter space.\\
\indent We also observed that RID allows convenient control over the optical loss, primarily through reflectivity constraints imposed on individual components. A detailed characterization was performed by parameter sweeping an HS-PIC with identical reflective nodes. As shown in Fig.~\ref{fig2}G, by setting appropriate limits on reflectivity and recursion order, the transmission power of the network can be maintained above $-5$\,dB.\\
\indent To demonstrate the unparalleled spectral manipulation capabilities, we applied our recursive inverse design framework to a single waveguide structure under experimentally measured dispersion and propagation loss of the waveguides. The design was implemented on a 300\,nm single-layer silicon nitride-on-insulator (SiNOI) platform to leverage its low optical loss, which increases the participation of multiple reflected photonic paths and thereby enriches the spectral response. Identical nodes are placed along the straight waveguide to form the simplest chain-structured HS-PIC composed of only two-port routers. Optical loss constraints are set to be 5\,dB (colored data points in Fig.~\ref{fig2}G show the corresponding numerical values). Across five complexity targets and nineteen operating spectral bands, the process yielded 95 distinct sets of structural parameters $\theta$ for the waveguide networks. These parameters are composed of sub-spaces ($\theta^{(i)}$), each of which represents either a router or a waveguide segment, as represented in Fig.~\ref{fig2}A-E green vectors. The result shows that all 95 circuits achieved the specified design goals in simulation. The optimized routers naturally exhibit a chirped-grating-like profile with near-zero intrinsic loss (Fig.~\ref{fig2}A right). Within sixteen routers and a waveguide length of 4.5\,cm, the waveguide spanned a complexity design space of more than $10^4$\,times, achieving a complexity value up to 111\,channels\,per\,nm (Fig.~\ref{fig2}B-E, right). With a fabrication-compatible feature size of 250\,nm and an ultra-compact device footprint of $30\,\mu m^{2}$, the router operates from visible to mid-infrared, covering 760 to 2100\,nm.\\
%
%Fig3
\begin{figure*}[ht]
\centering
\includegraphics[width = 18cm]{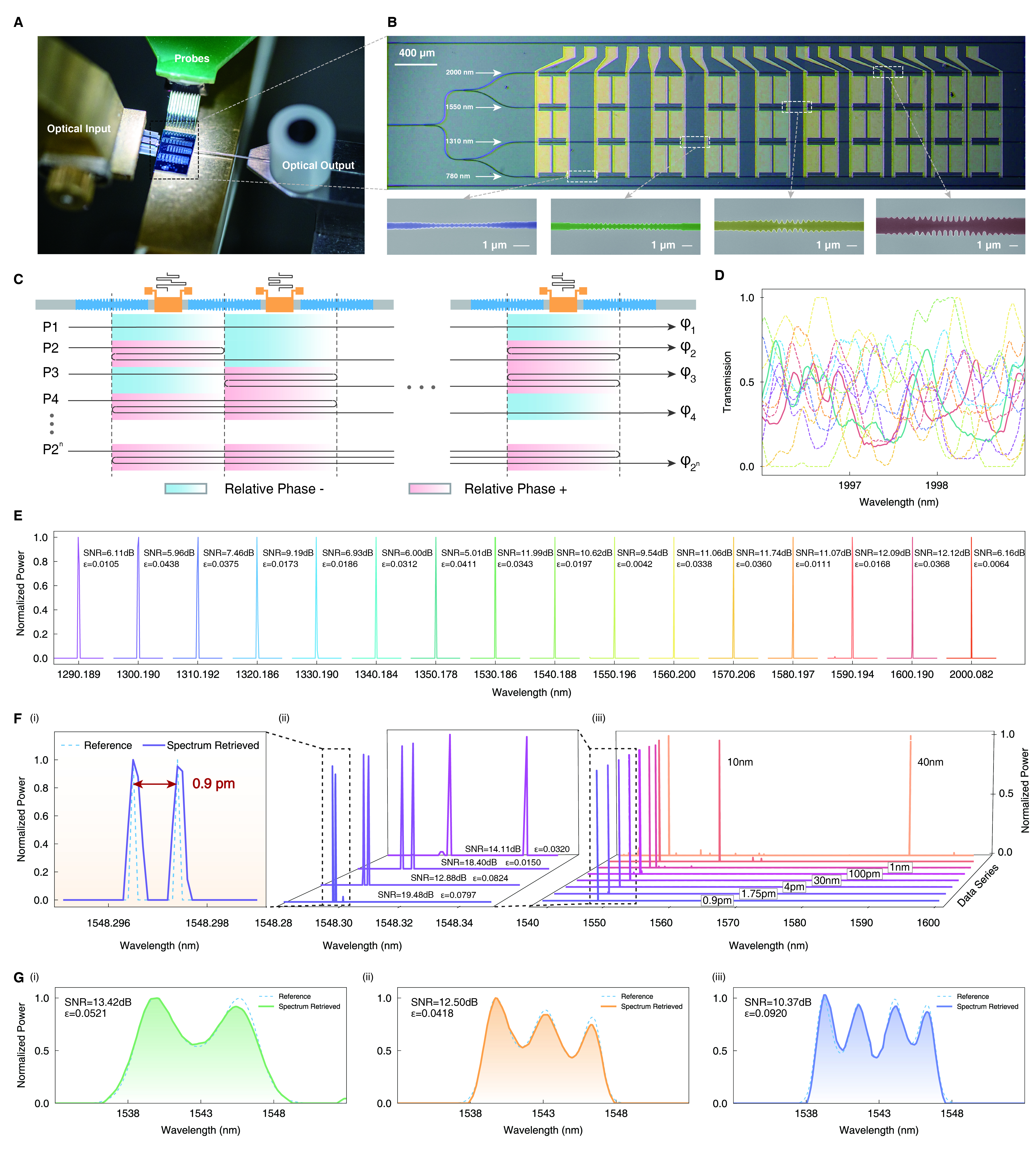}
\caption{\textbf{Setup and Testing of the Ultra-resolution Spectrometer.} ({\bf A}) Testing setup of the spectrometer (left), ({\bf B}) Optical charged couple device (CCD) image of the fabricated chip (upper), and fake color scanning electron microscope (SEM) of the routers at different operation bands (lower). ({\bf C}) The working principle of the spectrometer where phase shifters modulate the relative phase of optical pathways. ({\bf D}) Sample channels of the Spectrometer, dashed lines are randomly selected channels and the solid lines are two closest channels, showing a significantly different response. Reconstruction result of ({\bf E}) Single-peak laser ({\bf F}) dual-peak laser, and ({\bf G}) wide-band laser. The noise and performance are marked by SNR and $\varepsilon$.}
\label{fig3}
\end{figure*}

%Result Part3
\vspace{3pt}
\noindent\textbf{Hyper-resolution spectrometer based on HS-PICs}\\
\noindent Our HS-PIC enables ultrahigh-resolution spectral sensing. Fig.~\ref{fig3}C outlines the operating principle when a thermally tuned hyper-spectral waveguide is used as a spectrometer. The spectrum modulation at the output comes from the interference of all propagation pathways induced by different reflection positions. Only paths with zero or twice the reflections are counted because higher-order reflections have a negligible power proportion. Thermal tuning of the waveguides introduces an effective index change $\Delta n$, which produces controlled phase separation between the single-pass and triple-pass routes. Heating combinations across segments therefore set the relative phases of paths and produce a family of distinct spectral responses. Experimentally, we preset three discrete voltage states (0\,V, 3.5\,V and 5\,V) for each HS-PIC waveguide that exhibit sufficiently low pairwise spectral correlation (Fig.\,S4). In this setup, an 11-stage HS-PIC waveguide operates as a spectrometer with 59,049 mutually independent response channels.\\
\indent The experimental configuration of the HS-PIC spectrometer is presented in Fig.~\ref{fig3}A. The system is composed of a silicon nitride HS-PIC chip fabricated on a 4-inch, 300\,nm SiN wafer (see Supplementary Material, Method 3 for chip fabrication), optical edge couplers, a direct current (DC) probe card, and a high-speed driving board with a microcontroller unit (MCU). The chip routes input light via three Y-branches into four HS-PIC waveguides, each targeting a different spectral band (780, 1310, 1550, and 2000\,nm) for parallel spectrum detection. Under passive and biased operation, the chip exhibits a low comprehensive loss of 16\,dB, 14\,dB, 11\,dB, and\,13 dB for four ports under corresponding ASE light sources, while blocking the light of other bands. Calibration was performed with four different ASE sources because of the limited bandwidth, activating a total spectral coverage of 235\,nm. The maximum operation bandwidth is limited by the ASE source, which can be inferred from the symmetry of the response envelope to surpass 400\,nm (Supplementary Material, fig.S5). The maximum power consumption of each waveguide is experimentally measured around 49\,mW, resulting in an average system power consumption of 245\,mW. The HS-PIC chip has a fast channel switching time below 0.5\,ms (Supplementary Material, fig.\,S7), supporting real-time spectral acquisition. In its normal operating mode with 1024\,channels, the HS-PIC performed a scan in 1\,s, with channel count and acquisition time increasing under low-SNR conditions.\\
\indent To characterize the HS-PIC performance, we performed spectral reconstruction of single-peak, dual-peak, and wide-band lasers by scanning the intensity response of all channels. The spectral encoding of our HS-PIC spectrometer represents an underdetermined system, which we solved using a least mean squares (LMS) method with regularization constraints. The norm terms in our recovery algorithm are expressed as:
\begin{equation}\min_{0<\phi< 1}\|I-T\phi\|_{2}+\alpha\|\phi\|_{1}+\beta\|\phi\|_{2}+\gamma\|D\cdot\phi\|_{1}\end{equation}
Where $\alpha$, $\beta$, $\gamma$ denote the coefficient of the L1, L2 and total variation (TV) norms\cite{yao2023Integrated}. To quantify errors in the signal acquisition and processing, we defined the signal-to-noise ratio (SNR) of the detected signal and the spectral recovery accuracy ($/eta$), which are given by $\mathrm{SNR}_{\mathrm{dB}}=10\log_{10}(\|I\|_{2}^{2}/\|I-I_{0}\|_{2}^{2})$ and $\varepsilon=\left|\left|\boldsymbol{\phi}_0-\boldsymbol{\phi}\right|\right|_2/\left|\left|\boldsymbol{\phi}_0\right|\right|_2$, respectively. The $I$ and $I_{0}$ represent the detected and ideal intensity response. Results for single-peak lasers (Fig.~\ref{fig3}E) demonstrate excellent uniformity and accuracy across the operational bandwidth ($\varepsilon < 0.05$). Fig.~\ref{fig3}F presents the spectral results for dual-peak lasers with various frequency spacings. For larger spacings (4\,pm to 40\,nm), the input light was generated by two independent lasers, while for small spacings (1.75\,pm and 0.9\,pm), it was produced by a laser modulated by one or two 100MHz acousto-optic modulators (AOM). This modulation approach is necessary to generate a stable dual-peak signal in the presence of approximately 10\,MHz laser frequency jitter. The resolution might be pushed further with a more stable laser. Simulations further suggest that, with an ideal jitter-free source, the achievable resolution could be improved to approximately 0.1\,pm. For wide-band light (Fig.~\ref{fig3}G), while the SNR was reduced compared with dual-peak lights, the high channel number of 4096 provided sufficient redundancy to maintain a low recovery error ($\varepsilon < 0.1$).\\
\indent The spectrometer performance is limited by dynamic errors during the detection process. First, temperature drift from phase shifters limits the ultimate resolution. We experimentally verify the chip having a thermal sensitivity of 0.014\,pm/mK (see Supplementary Material, note 2, for detailed experiment results and analysis). To mitigate this, we optimized our channel scanning strategy to minimize power fluctuation between adjacent phase shifters. This approach successfully reduced the on-chip temperature drift by an order of magnitude to less than 10\,mK, which corresponds to a spectral accuracy of 0.14\,pm. This provides the engineering foundation for our sub-picometer, high-precision resolution. Second, HS-PIC spectrometers demonstrate strong robustness against power fluctuations. The unpackaged in-lab testing induces about 1 dB error on intensity response ($I$), proven by low SNR in most experiments. However, the high number of scanning channels provides sufficient redundant information, allowing signal recovery with good fidelity even at a low SNR ratio. This resilience also suggests the spectrometer's potential for robust operation in low-light environments.\\
%Fig4
\begin{figure*}[ht]
\centering
\includegraphics[width = 18cm]{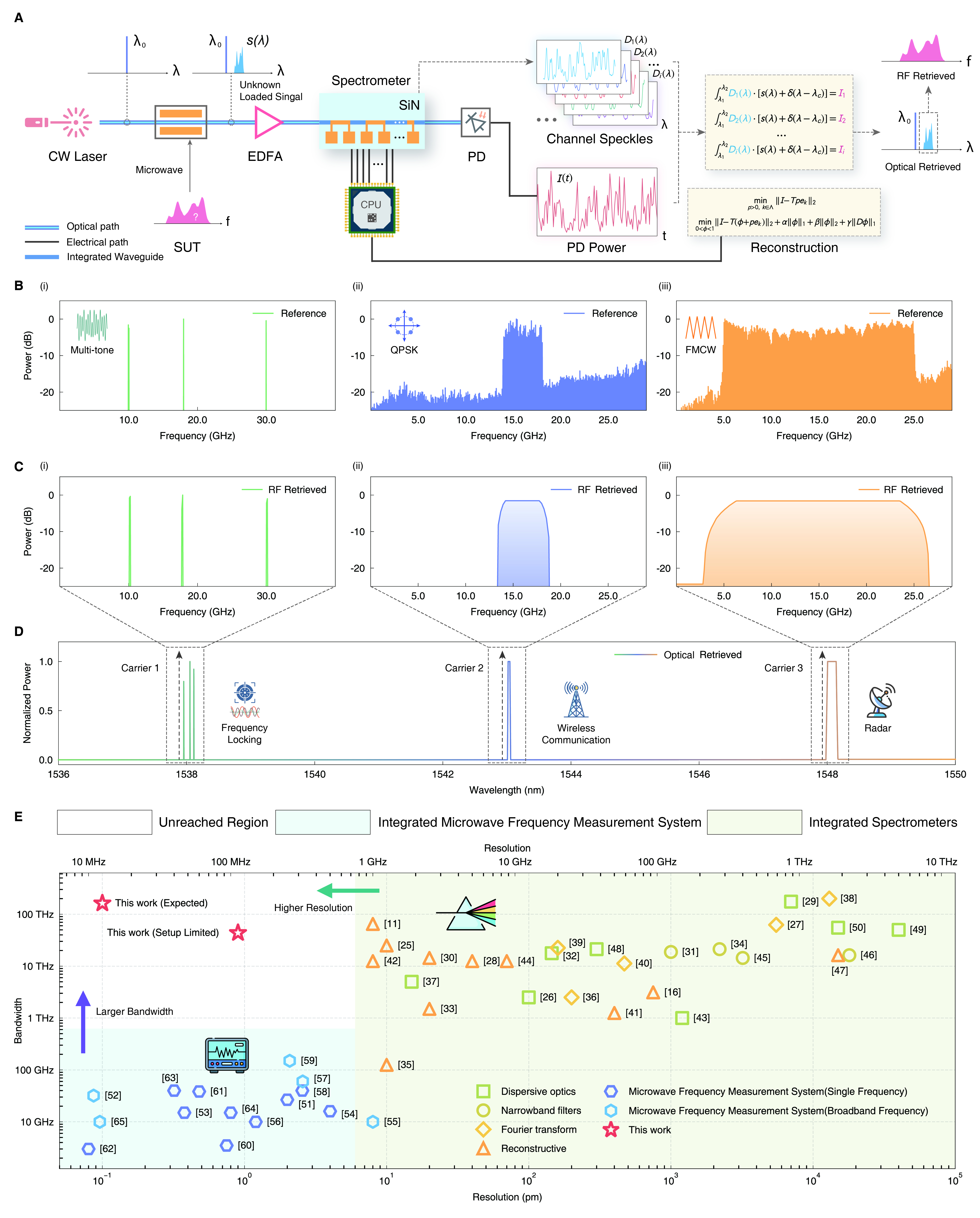}
\caption{\textbf{Bridging the Gap Between Optical and Microwave Spectrum Analysis.} ({\bf A}) Experimental setup of the HS-PIC microwave analysis system (left) and its reconstruction steps (right). The pink and blue signals correspond to the microwave and optical signals, respectively.  ({\bf B-D}) Comparison between the detection system and electrical spectrum analyzer (ESA) for (i) multi-tone, (ii) QPSK, and (iii) FMCW signal recovery.
(B) Results measured by ESA. (C) Results obtained from the detection system. (D) Intermediate optical-domain recovery states for different carrier frequencies. ({\bf E}) A collection of recent works on spectrum analysis shows a gap between optical and microwave detection.}
\label{fig4}
\end{figure*}

%Result Part4
\vspace{3pt}
\noindent\textbf{Unified microwave-optical spectrum analyzing}\\
\noindent The hyper-resolution accuracy of our approach unlocks previously unattainable opportunities for on-chip applications. As recent advances are summarized in Fig.~\ref{fig4}E, integrated spectrometers work in several hundred nanometers bandwidth yet are constrained to near 10 pm maximum resolutions\cite{cheben2007Highresolutiona,pohl2020Integrateda,xu2023Breakinga,cheng2019Broadbanda,yao2023Broadbanda,nitkowski2008Cavityenhanceda,yao2024Chipscale,koshelev2014Combinationa,zhang2021Compacta,redding2013Compact,emadi2012Designa,redding2016Evanescentlya,kita2018Highperformancea,lin2023Highperformancea,grotevent2023Integrated,yao2023Integrated,li2021Integrateda,zheng2019Microringa,zhao2024Miniaturizeda,zhang2025Miniaturizeda,momeni2009Planara,zhang2025Scalablea,ryckeboer2013SilicononInsulatora,correia2000Singlechipa,yang2019Singlenanowirea,zhu2017Ultracompacta,sander1996Optical, web1}, while on-chip microwave frequency analyzers attains a high resolution of about 100\,MHz within only about 1\,nm of equivalent optical bandwidth\cite{jiao2021Demonstrationa,tao2022Fullya,song2021Highresolutiona,zhu2019HighSensitivitya,yao2022Highlya,liu2020Instantaneousa,wang2024Integrateda,pagani2015Lowerrora,shi2025Microwavea,marpaung2013OnChipa,chen2019Onchipa,zhao2023PhotonicAssisteda,wang2025PowerIndependenta,huang2025Simplea,ding2023Simultaneousa}. Using the same chip described in Section 3, we demonstrate a microwave spectrum analyzer that bridges this performance gap, achieving both wide bandwidth and high resolution on the same platform. As the experiment setup in Fig.~\ref{fig4}A, left panel, the signal-under-test (SUT) is loaded using a carrier-suppressed single-sideband (CS-SSB) lithium niobium (LN) modulator, then encoded by our chip encoder. The Reconstruction process of the microwave follows the steps in Fig.~\ref{fig4}A, right panel. Because of the presence of a strong reference carrier, the retrieval algorithm is adapted to a two-step LMS procedure, allowing for more accurate recovery of both the carrier and signal, with the two norms separately written as:
\begin{equation}\min_{p>0,k\in \Lambda}\|I-Tpe_{k}\|_{2}\end{equation}
\begin{equation}\min_{0<\phi< 1}\|I-T(\phi+pe_{k})\|_{2}+\alpha\|\phi\|_{1}+\beta\|\phi\|_{2}+\gamma\|D\phi\|_{1}\end{equation}
where $T$ is the transmission matrix from calibration. $pe_{k}$ and $\phi$ denote the carrier and signal within reconstruction window $\Lambda$ respectively. We demonstrate the reconstruction on multiple carrier frequencies and signal formats to validate its generality. As shown in Fig.~\ref{fig4}B-D, the multi-tones, QPSK, and FMCW inputs are recovered with a high resolution of 125\,MHz. The average frequency mismatch, defined as the mean error across all center and edge frequencies, was calculated to be 157\,MHz. It is slightly larger than the 0.9\,pm spectrometer resolution, which is attributed to carrier-induced intensity noise. The HS-PIC thus delivers high-resolution for both microwave and optical spectra and offers a competitive route to unified on-chip frequency domain analysis.\\

% Discussions
\vspace{3pt}
\noindent\textbf{Discussions} \\
\noindent Our HS-PIC architecture combines high spectral complexity with extensive tunability, enabling spectral control with unparalleled versatility. This capability unlocks new possibilities across multiple photonic applications. In optical communications, it allows aggregate multichannel modulation of parallel sources within a single waveguide, even for combs with free spectral ranges (FSRs) below 100\,GHz, thus making parallel transmission as straightforward as single-wavelength operation and eliminating the need for bulky separation–recombination components such as AWGs or MUX/DeMUX modules\cite{zhang2024Highcoherence}. For spectroscopic sensing, sub-GHz resolution has long depended on optical–electronic hybrid sampling techniques that suffer from poor stability and face significant integration challenges\cite{picque2019Frequency,xu2024Nearultraviolet, han2024Dualcomb,yang2022Multidimensional}. HS-PIC provides an all-optical alternative, offering compactness, stability, and scalability. In optical computing, the HS-PIC enables highly parallel and tunable processing cores by co-designing multi-channel composite responses, offering a promising route toward general-purpose photonic processors that reconcile tunability and channel density within a compact footprint. Future integration with advanced routers and low-loss electro-optic phase modulators may further extend its operational bandwidth and switching speed, enabling hyper-spectral cross-band GHz control within a single waveguide.\\
\indent Beyond these applications, the RID framework underpins this progress by providing a generalizable strategy that unifies device physics, system functionality, and design automation. While applied here to optimize the HS-PIC, RID naturally extends to broader photonic platforms. It transforms the traditionally manual, intuition-driven design process into a data-guided and algorithmically automated workflow. This approach brings photonic design closer to the automation and abstraction levels of modern electronic design automation (EDA). Such system-level automation could greatly accelerate the development and deployment of complex photonic integrated circuits, paving the way for widespread adoption of PICs across computing, communication, and sensing platforms.\\

\vspace{3pt}
\begin{footnotesize}
\noindent\textbf{Data availability}\\
%The authors declare that the data supporting the findings of this study are available within the article and its Supplementary Information. All raw data are available from the corresponding author upon reasonable request.
The data that support the plots within this paper and other findings of this study are available from the corresponding authors upon reasonable request. 

\vspace{3pt}
\noindent\textbf{Code availability}\\
The codes that support the findings of this study are available from the corresponding authors upon reasonable request.

\end{footnotesize}
\vspace{20pt}

%_______REFERENCE____________%
\bibliography{REF.bib}
% \bibliographystyle{naturemag}
%\bibliographystyle{naturesaa}

%%_______REFERENCE_______%%

%--------------------------------------------------------------------------
\vspace{12pt}
\begin{footnotesize}

\vspace{6pt}
\noindent \textbf{Acknowledgment}

\noindent The authors thank Ruoyu Shen for discussion of the initial concept, thank Junqi Wang, Haoyang Luo, Xin Zhou, Yinke Cheng, Ruixiang Song and Yiqing Liu for fabrication help, thank Zuoqi Zhong and Zixuan Zhou for valuable comments, thank Shenzhen Golight Technology Co., Ltd and the Institute of Semiconductors, Chinese Academy of Science for providing key equipment. L.C. acknowledges the Beijing Municipal Natural Science Foundation (Grant No. Z220008), the National Key Research and Development Program of China (Grant No. 2021YFB2801200), the Beijing Municipal Science $\&$ Technology Commission, Administrative Commission of Zhongguancun Science Park (Grant No. Z231100006023007), the National Natural Science Foundation of China (Grant No. 12293052), Joint Research Project of the Shijiazhuang-Peking University Cooperation Program. This work is supported by the High-performance Computing Platform of Peking University and Advanced Photonic Integrated Center of Peking University.

\vspace{6pt}
\noindent \textbf{Author contributions}

\noindent 
The concept of this work was conceived by H. H. and L. C. The HS-PIC theory and RID design were developed by H. H., L. C and Z. T. The chip was fabricated by Y. W., H. H. and H. Z. and Q. Y. The experiments were performed by H. H. and Z. T., with the assistance of C. F., Y. Z., Y. C. and R. Y. The reconstruction algorithms were developed by H. H. and Z. T. All authors participated in writing the paper. The project was under the supervision of L.C.
\end{footnotesize}
%--------------------------------------------------------------------------

\end{document}